%% file: main.tex
\documentclass[journal]{IEEEtran}

\usepackage[utf8]{inputenc}
\usepackage{amsmath}
\usepackage{cite}
\usepackage{multicol}
\usepackage{graphicx}
\usepackage{array}
\usepackage{multirow}
\usepackage{graphicx}
\usepackage{enumitem}
\usepackage{color}
\usepackage{nomencl}
\usepackage{etoolbox}
\usepackage{amssymb}
\usepackage{comment}
\usepackage{mathtools}

\ifCLASSOPTIONcompsoc \usepackage[caption=false,font=normalsize,labelfon
t=sf,textfont=sf]{subfig}
\else
\usepackage[caption=false,font=footnotesize]{subfi g}
\fi

\usepackage{tikz}
\usetikzlibrary{decorations.pathreplacing}
\usetikzlibrary{arrows,shapes,positioning}
\usetikzlibrary{patterns}
\usepackage{pgfplots}
\pgfplotsset{compat=newest}
\pgfplotsset{plot coordinates/math parser=false}
\newlength\figureheight
\newlength\matlabfigurewidth
\newcommand\Item[1][]{%
  \ifx\relax#1\relax  \item \else \item[#1] \fi
  \abovedisplayskip=0pt\abovedisplayshortskip=0pt~\vspace*{-\baselineskip}}

\newcommand{\argmin}[1]{\underset{#1}{\operatorname{arg}\,\operatorname{min}}\;}

\title{An ADMM-based Coordination and Control Strategy for PV and Storage to Dispatch Stochastic Prosumers: Theory and Experimental Validation}
\author{
    \IEEEauthorblockN{
    Rahul Gupta$^1$,
    Fabrizio Sossan$^1$, Enrica Scolari$^1$, Emil Namor$^1$, Luca Fabietti$^2$, Colin Jones$^2$,
    Mario Paolone$^1$\\}
    \IEEEauthorblockA{$^1$Distributed Electrical Systems Laboratory, $^2$Automatic Control Laboratory\\
    EPFL, Lausanne, Switzerland
    \\}
    }
\usepackage{etoolbox}
\makeatletter
\patchcmd{\@maketitle}
  {\addvspace{0.5\baselineskip}\egroup}
  {\addvspace{-2\baselineskip}\egroup}
  {}
  {}
\makeatother
\begin{document}
\maketitle

\begin{abstract} 
This paper describes a two-layer control and coordination framework for distributed energy resources. The lower layer is a real-time model predictive control (MPC) executed at 10~s resolution to achieve fine tuning of a given energy set-point. The upper layer is a slower MPC coordination mechanism based on distributed optimization, and solved with the alternating direction method of multipliers (ADMM) at 5~minutes resolution. It is needed to coordinate the power flow among the controllable resources such that enough power is available in real-time to achieve a pre-established energy trajectory in the long term.
Although the formulation is generic, it is developed for the case of a battery system and a curtailable PV facility to dispatch stochastic prosumption according to a trajectory at 5~minutes resolution established the day before the operation.
The proposed method is experimentally validated in a real-life setup to dispatch the operation of a building with rooftop PV generation (i.e., 101~kW average load, 350~kW peak demand, 82~kW peak PV generation) by controlling a 560~kWh/720~kVA battery and a 13~kW peak curtailable PV facility.
\end{abstract}


\makenomenclature

\renewcommand\nomgroup[1]{%
  \item[\bfseries
  \ifstrequal{#1}{P}{\textit{Indices}}{%
  \ifstrequal{#1}{N}{\textit{Constants}}{%
  \ifstrequal{#1}{O}{\textit{Variables}}{}}}%
]}

\mbox{} 

\nomenclature[O]{$\mathcal{G}_i$}{Copied variable for PV set-point for $i$ interval [kW]}
\nomenclature[O]{$\mathcal{B}_i$}{Copied variable for BESS set-point for $i$ interval [kW]}
\nomenclature[O]{$y_{G_{i}}$}{Dual variable for PV problem for $i$ interval}
\nomenclature[O]{$y_{B_{i}}$}{Dual variable for BESS problem for $i$ interval}
\nomenclature[O]{$u_{G_{i}}$}{Scaled dual variable for PV problem for $i$ interval}
\nomenclature[O]{$u_{B_{i}}$}{Scaled dual variable for BESS problem for $i$ interval}
\nomenclature[O]{${G}_{i}$}{Original variable for PV set-point for $i$ interval [kW]}
\nomenclature[O]{$B_{i}$}{Original variable for BESS set-point for $i$ interval [kW]}
\nomenclature[O]{$SOC_{i}$}{BESS state of charge for $i$ interval [\%] }
\nomenclature[O]{$SOC_{i}$}{BESS state of charge for $i$ interval [\%] }
\nomenclature[N]{$SOC_i^\text{max}$}{Upper limit on BESS state of charge  for $i$ interval [\%]  }
\nomenclature[N]{$SOC_i^\text{min}$}{Lower limit on BESS state of charge  for $i$ interval [\%] }
\nomenclature[N]{$B^\text{max}$}{Upper limit of BESS actuation for [kW] }
\nomenclature[N]{$B^\text{min}$}{Lower limit on BESS actuation for [kW] }
\nomenclature[N]{$\widehat{G}_i$ }{Estimated PV MPP for $i$ interval [kW]}
\nomenclature[N]{$e_i$}{Tracking error at GCP for $i$ interval [kW]}
\nomenclature[N]{ $\widehat{e}_{i}$}{Forecasted Tracking error at GCP for $i$ interval [kW]}
\nomenclature[N]{$P_{disp}$}{Dispatched power [kW]}
\nomenclature[N]{$\hat{P}_{Load}$}{Forecast load [kW] }
\nomenclature[N]{$T_{s}$}{Sampling time for coordination mechanism}
\nomenclature[N]{$N$}{Length of prediction horizon }
\nomenclature[P]{$i$}{Index of 5 minute time interval from 1 to $N$}
\nomenclature[P]{$m$}{Index of 10 sec interval within each $i$ interval }
\nomenclature[P]{$k$}{Index of iterations within each $i$ interval}
\nomenclature[P]{$j$}{Rolling time index from $i$ to $N$}
\nomenclature[N]{$P$}{Realization at GCP [kW] }
\nomenclature[N]{$L$}{Load at GCP [kW]}
\nomenclature[N]{$\rho$}{Penalty parameter}
\nomenclature[N]{$E$}{Battery nominal energy capacity}




\section{Introduction}
Due to the displacement of conventional generation in favor of production from renewables, decentralized control schemes for distributed energy resources (DERs) have gained considerable attention in the recent literature to support grid operation.
Control frameworks for DERs have been focusing on achieving targets at both the local level, e.g., voltage control and congestion managements\cite{6598997, vovos2007centralized, hu2014coordinated}, and system level, e.g. primary  and secondary frequency regulation\cite{6558529, guerrero2013advanced, molina2011decentralized, liu2014decentralized}. Other reported methods  proposed open-loop control of DERs by broadcasting electricity price signals, as in \cite{Hammerstrom2007}, and virtual power plant strategies to enable control by explicit power set-points\cite{8013070, Bernstein2015}. A concept which has emerged in the recent literature is dispatching the active power flows of stochastic resources according to a trajectory established before the operation thanks to controlling energy storage systems, as in \cite{7542590, 6913566}, in combination with flexible demand\cite{fabietti2017j, saele2011demand, borenstein2002dynamic, sundstrom2012flexible}, and also accounting for grid constraints \cite{7948761}.

The advantage of dispatching traditional stochastic resources is that it inherently reduces the amount of regulating power required to operate the grid without the necessity of coordination between the local system and grid operators. With respect to control strategies for virtual power plants, which normally consists in myopically tracking a power set-point, the dispatch problem as defined in \cite{7542590} implicitly includes an energy management policy that does represent a key feature in the context of managing the flexibility of energy storage systems.

In this paper, we describe the formulation and experimental validation of a control framework to coordinate the operation of heterogeneous DERs to dispatch the active power flow at the grid connection point (GCP) of an active distribution network (ADN) according to a trajectory, called dispatch plan, determined the day before the operation. The framework consists in two algorithmic layers running at different paces. The lower layer, executed at 10~s resolution, has the objective of pursuing an accurate tracking of the dispatch plan at the GCP. The upper layer, executed once each 5~minutes and implemented through distributed optimization, has the objective of coordinating available neighboring DERs such that enough power capacity is available in real-time to achieve a successful dispatch. In the existing literature, distributed optimization has been applied to meet manifold objectives such as solving optimal power flow problems \cite{erseghe2014distributed}, optimal dispatch of photovoltaic (PV) inverters in residential distribution networks \cite{dall2014decentralized} and  demand response strategy with electric vehicles (EVs) \cite{tan2014optimal}. Unlike centralized methods, it is highly scalable, an appealing feature in the context of rapidly growing distributed energy networks, and it exhibits better privacy and security properties compared to centralized schemes. Controlling and coordinating multiple units in real-time can be effectively achieved through distributed methods, thanks to the Douglas-Rachford splitting \cite{giselsson2017linear} which divides the problem into small objectives and solve it through an aggregator. For distributed methods, on the contrary to dual decomposition that employs the subgradient projection technique with quite slow convergence, the Douglas-Rachford splitting technique is faster \cite{7419888} and efficient even when applied to non-linear objectives.
In this paper, the alternating direction method of multipliers (ADMM)\cite{boyd2011distributed}, a special case of Douglas-Rachford splitting, is applied to solve the distributed optimization problem. The proposed algorithmic toolchain is tested in a real-life experimental setup to dispatch the operation of a group of stochastic prosumers by controlling a battery energy storage system (BESS) and a curtailable rooftop PV facility.

A similar approach was proposed in \cite{fabietti2017j} to control the operation of a BESS and a building with flexible demand (electric space heating). With respect to \cite{fabietti2017j}, in this paper we apply ADMM, which represents more agnostic formulation since all DERs share the same information.

The paper is organized as follows: Section~\ref{sec:problem_stat} states the problem we intend to solve, Section~\ref{sec:methods} describes the formulation of the proposed framework, Section~\ref{sec:results} presents the experimental results and Section~\ref{sec:concs} summarizes the contributions of this paper and states the conclusions.

\section{Problem Statement}\label{sec:problem_stat}
The objective of the proposed control strategy is to dispatch the active power flow of a distribution network according to a deterministic profile, called \emph{dispatch plan}, established the day before the operation. The dispatch plan is a sequence of average power flow values at 5 minute resolution. It is denoted by $P_{disp}$ and, in this work, it is assumed known and computed with the procedure described in \cite{7542590}. In brief, the dispatch plan consists in the sum of two terms: the forecast of the prosumption underneath the distribution system (i.e., intuitively, it is the best guess one could do to dispatch stochastic prosumption), and the offset profile which is with the specific objective of restoring a suitable level of flexibility in the available storage resources\footnote{For example, during the current day of operation, flexible resources might be close to their state-of-charge bounds. Therefore, it is necessary to charge/discharge them to restore an adequate level of SOC such that enough flexibility is available to compensate for the tracking error during the incoming day of operation.}. It is noteworthy that in the current formulation, we do not include grid constraints (e.g., see \cite{7948761, 8013070}), which will be considered in future works. In other words, we assume that the target grid, as the one considered in the real-life experimental setup, is robust and not subject to voltage and line ampacity constraints violations.

\begin{figure}[!ht]
    \small{
    \centering
    \input{fig/network_diagram.tex}
    }
    \caption{The EPFL's experimental setup used for validation: a MV feeder interfacing office buildings equipped with uncontrollable rooftop PV plants, a grid-connected battery system, and a curtailable PV plant.}
    \label{fig:networkdiagram}
\end{figure}
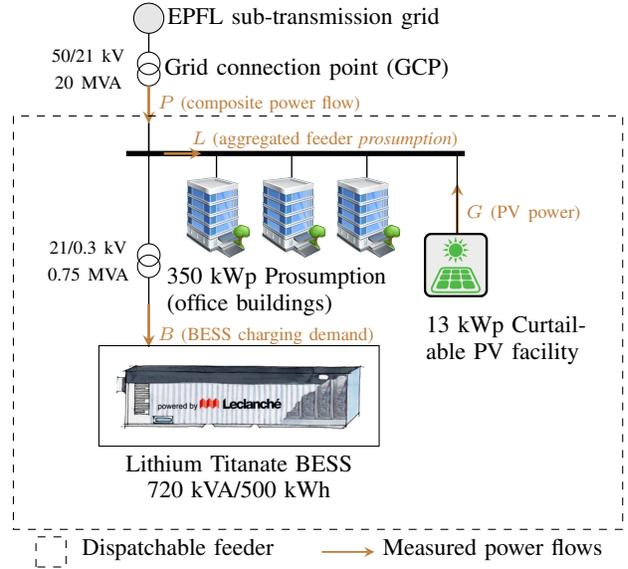

In real-time, the problem is to compensate the average power mismatch between the dispatch plan set-point and the measured stochastic realization, on a 5 minute basis. This is achieved by the coordinated control of multiple resources thanks to applying distributed optimization. Although the proposed formulation is generic enough to accommodate the use of different kinds of flexible resources, in this paper it is formulated for the case of a curtailable PV facility and battery energy storage system (BESS). The choice of these kinds of units is motivated by the fact that they are both well-established technologies, characterized by a high level of maturity and foreseen to undergo a massive growth in the near future. Fig.~\ref{fig:networkdiagram} shows the real-life experimental setup used to test and validate the proposed control: it consists in a radial distribution feeder interfacing a heterogeneous mix of demand and generation (office buildings with uncontrollable rooftop PV installations) and two controllable resources as explained in Section~\ref{Expt. setup}.

\section{Methods}\label{sec:methods}
The objective is to coordinate the operation of and control multiple resources to compensate the mismatch between the dispatch plan set-point and the measured average power consumption on a 5 minute basis. In our case, the formulation is carried out by considering a curtailable PV facility and a BESS because they are the unit available in the experimental setup used for the real-life validation. Nonetheless, the setup is scalable and can be augmented to consider other controllable resources. The control framework consists in two real-time algorithms executed at different paces:
\begin{itemize}
\item a lower level tracking problem, executed at 10~s resolution, to achieve a fine tracking of the dispatch plan by controlling the BESS's active power set-point. This is achieved with MPC with the formulation proposed in \cite{7542590}, and it is not a contribution of this work.
\item an upper level coordination mechanism, running at 5~minutes resolution, to coordinate the operation between elements. Its role is essentially implementing an energy management strategy to, \emph{i)}, make sure that enough power capacity is available for the faster control loop to compensate for the power mismatch, and, \emph{ii)}, longer-term managing the BESS state-of-charge (SOC) such that enough flexibility is available to compensate for the energy error during the remaining part of the day.
\end{itemize}
The latter control loop is the main contribution of this paper and consists in an optimization problem formulated and solved for each controllable device with a coupling constraint on the energy error. In the following formulation, the upper level problem is responsible for computing set-points for both the PV plant and battery system. 

It is worth noting that decoupling the control actions into two discerned time scales is practical for two reasons: \emph{i)}, the distributed optimization coordination mechanism is slower to be solved than the faster pace power set-point tracking (especially because of the communication latency when agents are implemented on different machines); \emph{ii)}, due to technological limitations or inherent process uncertainties, certain devices (like PV plants and fuel cells) are not curtailable/controllable with high accuracy, see e.g.\cite{scolari_2018}. Therefore, their use is more likely for energy management rather than precise power point tracking. On the other hand, resources like grid-connected battery systems are able to achieve accurate power control and are suitable for fine power point tracking. Overall, the proposed solution (i.e., fast pace control coupled with slower distributed optimization problem) allows accommodating both kinds of resources in a cooperative and thoughtful manner.

In the following of this section, the model predictive control problem is first formulated in a centralized way. Later in this section, it is shown how it is decomposed until having one problem per each controllable unit, and how it is solved by applying the ADMM algorithm.

\subsection{Centralized model predictive control}
Let the index $i$ denote the rolling current 5 minute interval, $N$ the number of 5~minutes interval in 24 hours, $j=i, \dots, N$ a 5 minute time index spanning from the current time until the end of the day, the sequence $\widehat{e}_{j}$ the forecasted deviation between the dispatch plan set-point and forecasted realization for the remaining part of the day, $B^o_j$ and $G^o_j$ the battery and PV plant set-point trajectories, respectively, and $\widehat{G}_j$ the maximum theoretical PV power plant output. The sequences $\widehat{e}_{j}$ and $\widehat{G}_j$ are calculated by applying forecasting tools and are assumed given in the following formulation. For our specific application, the latter is determined with the physical-based modelling tool-chain proposed in \cite{scolari2017}, as a function of global horizontal irradiance forecast. In the following, we focus on the formulation of the optimization problem in a centralized manner. It is aimed at determining trajectories $B^o_j$ and $G^o_j$  for the remaining part of the day $j=i\dots, N$  in order to compensate for the dispatch plan mismatch $e_j$ while performing minimal amount of curtailment subject to PV and battery system constraints. The PV system constraint is that the active power should be within 0 and the theoretical maximum PV power production $\widehat{G}_j$. For the battery, the constraints are that the power should be within the four-quadrant apparent power converter capability, and the battery SOC within the bounds ($\text{SOC}_j^\text{min}, \text{SOC}_j^\text{max}$) which can be time variant to account for dynamic constraints. In this formulation and experiments, the battery converter is operated at the unitary power factor, so the apparent power constraint reduces to the active power being between plus/minus the converter apparent power rating\footnote{Power converter capability depend on grid frequency and voltage, in this case we choose a conservative $B^\text{nom}$ bound so to satisfy any operating condition.}, denoted by $B^\text{nom}$. Battery SOC is normally modelled considering the charging/discharging efficiency of the system, nevertheless, in the current formulation we neglect it because the considered experimental unit has a round-trip efficiency on the AC side close the unit (in the range 97-99\% according to the power)\footnote{We note that one might model battery efficiency while still preserving convexity, see, e.g., \cite{7542590}. Another strategy to account for the efficiency is to account for efficiency losses in the next rescheduling interval.}. The dynamic model of the battery's SOC is:
\begin{align}\label{eq:socmodel}
\text{SOC}_{j+1} = \text{SOC}_{j} + \alpha B_j, ~~~ \alpha = \frac{300}{3600}\frac{1}{E} 
\end{align}
where the ratio 300/3600 is the share of 5~minutes period in an hour, and $E$ the battery nominal energy capacity in kWh. The SOC value at time interval $i+1$ is calculated by applying~\eqref{eq:socmodel} recursively to propagate the current battery state-of-charge $\text{SOC}_i$:
\begin{align}
\text{SOC}_{i+1} &= \text{SOC}_i + \alpha B_{i} + \dots + \alpha B_j = \\
&= f(\text{SOC}_i, \mathbf{B}_{i,j}) \label{eq:soc},
\end{align}
where the bold typeface notation $\mathbf{B}_{i,j}$ denotes the sequence $[B_i, \dots, B_{j}]$.

Formally, the centralized optimization problem is:
\begin{align}
\argmin{
\begin{matrix}
\scriptstyle G_{i}, \dots, G_{N}\\
\scriptstyle B_{i}, \dots, B_{N}
\end{matrix}
}  \sum_{j=i}^N \left( G_j-\widehat{G}_j \right)^2 \label{eq:cent:cost} 
\end{align}
subject to:
\begin{align}
& B_j + G_j  = e_j && j=i, \dots, N \label{eq:cent:c1} \\
& 0 \leq G_{j} \leq  \widehat{G}_j && j=i, \dots, N \label{eq:cent:c2}\\
& B^\text{min} \le B_{j} \le  B^\text{max} && j=i, \dots, N \label{eq:cent:c3}\\
& \text{SOC}^\text{min}_j \le f(\text{SOC}_i, \mathbf{B}_{i-1,j-1}) \le \text{SOC}^\text{max}_j && j=i, \dots, N\label{eq:cent:c4}.
\end{align}
In words, it determines the PV trajectory at minimum curtailment and the BESS power trajectory respectful of battery's constraints to satisfy the tracking error on the shrinking horizon from $i$ to $N$.
The optimization problem \eqref{eq:cent:cost}-\eqref{eq:cent:c4} is centralized because, to be solved, it requires all the information from the controllable resources. By exploiting well-known results from distributed optimization theory, it is possible to decompose the centralized problem into smaller ones, which can be solved iteratively until reaching a consensus on a coupling constraint -- in this case, the dispatch constraint \eqref{eq:cent:c1}.

\subsection{From the centralized to distributed optimization problem and ADMM}
Let $g_j(\cdot)$ be a barrier function with zero cost when the tracking error constraint~(\ref{eq:cent:c1})  is respected and infinity otherwise: 
\begin{align}\label{eq:barrier}
 g_j(G_j, B_j) =
\begin{cases} 
    0 & B_j + G_j  = e_j \\
    \infty & \text{otherwise.}
\end{cases}
\end{align}
 Let $\mathcal{G}_j$, $\mathcal{B}_j$ be the variables that mimic the behavior of the original variables ${G}_j$, ${B}_j$, the so-called copied variables. The centralized objective can be re-written by moving the system constraints into the objective function by using the barrier function $g_j(\cdot)$. It is:
\begin{align}
&\begin{aligned}
\argmin{
\begin{matrix}
\scriptstyle G_{i}, \dots, G_{N}\\
\scriptstyle B_{i}, \dots, B_{N}
\end{matrix}
}  \sum_{j=i}^N \left( G_j-\widehat{G}_j \right)^2  
& + \sum_{j=i}^N g_j(\mathcal{G}_{j}, \mathcal{B}_{j}) 
\end{aligned}\label{distributed}
\end{align}
subject to:
\begin{align}
& B_j - \mathcal{B}_j  =0 && j=i, \dots, N \label{eq:B_copy} \\
& G_j - \mathcal{G}_j  =0 && j=i, \dots, N. \label{eq:G_copy} 
\end{align}
Constraints \eqref{eq:B_copy} and \eqref{eq:G_copy} can be moved in the cost function by using two sequences of Lagrangian multipliers, denoted by $\boldsymbol{y_{G}}_{i,N}$ and $\boldsymbol{y_{B}}_{i,N}$, referred to as dual variables in the following.
The augmented Lagrangian cost function is obtained by moving the equality constraints \eqref{eq:B_copy}-\eqref{eq:G_copy} in the cost function \eqref{distributed}. It is:
\begin{align}
&\begin{aligned}
L_{\rho}=
& \sum_{j=i}^N \left( G_j-\widehat{G}_j \right)^2 + 
\sum_{j=i}^N g_j(\mathcal{G}_{j}+\mathcal{B}_{j}) + \\
+ &\frac{\rho}{2} \left( \left|\left| \boldsymbol{G}_{i,N} - \boldsymbol{\mathcal{G}}_{i,N} \right|\right|_2^2 
+ \left|\left| \boldsymbol{B}_{i,N} - \boldsymbol{\mathcal{B}}_{i,N}  \right|\right|_2^2 \right) +  \\
+ &\boldsymbol{y_{G}}^T_{i,N}(\boldsymbol{G}_{i,N} - \boldsymbol{\mathcal{G}}_{i,N}) + \boldsymbol{y_{B}}^T_{i,N}(\boldsymbol{B}_{i,N}- \boldsymbol{\mathcal{B}}_{i,N}).
\end{aligned}
\end{align}
The ADMM consensus and sharing problem \cite{boyd2011distributed} consists in the following three steps.

\subsubsection{Original variables updates}
Let  $\boldsymbol{u_{G}}_{i,N}$ =  $\boldsymbol{y_{G}}_{i,N}/\rho$ and $\boldsymbol{u_{B}}_{i,N}$ = $\boldsymbol{y_{B}}_{i,N}/\rho$ be the scaled dual variable for the PV and battery problems. The update of the decision variable of the PV power plant  is:
\begin{align}
&\begin{aligned}
\boldsymbol{G}_{i,N}^{k+1}  = \argmin{G_{i}, \dots, G_{N}} \Bigg\{\sum_{j=i}^N \left( G_j-\widehat{G}_j \right)^2  + \\ + \frac{{\rho}^{k}}{2} \left|\left| \boldsymbol{G}_{i,N} - \boldsymbol{\mathcal{G}}^k_{i,N} + \boldsymbol{u_G}_{i,N}^k \right|\right|_2^2  \Bigg\},
\label{OriginalUpdatePV}
\end{aligned}
\end{align}
subject to \eqref{eq:cent:c2}. For the battery, it is:
\begin{align}
\boldsymbol{B}^{k+1}_{i,N} = \argmin{B_{i}, \dots, B_{N}} \Bigg\{ \frac{{\rho}^{k}}{2} \left|\left| \boldsymbol{B}_{i,N} - \boldsymbol{\mathcal{B}}^k_{i,N} + \boldsymbol{u_B}_{i,N}^k \right|\right|_2^2 \Bigg\}\label{OriginalUpdatebess},
\end{align}
subject to \eqref{eq:cent:c3} and \eqref{eq:cent:c4}.

\subsubsection{Copied variable update}
\begin{align}\label{eq:copiedvariable}
&\begin{aligned}
[\boldsymbol{\mathcal{G}}^{k+1}_{i,N}, \boldsymbol{\mathcal{B}}^{k+1}_{i,N}]  = \argmin{\mathcal{G}_{i}, \dots, \mathcal{G}_{N},  \mathcal{B}_i, \dots,  \mathcal{B}_{N}} \Bigg\{ \sum_{j=i}^N g_j(\mathcal{G}_{j}+\mathcal{B}_{j})\\ + \frac{{\rho}^{k}}{2} \left|\left| \boldsymbol{G}_{i,N}^{k+1}- \boldsymbol{\mathcal{G}}_{i,N} +  \boldsymbol{u_G}_{i,N}^k \right|\right|_2^2 + \\
\frac{{\rho}^{k}}{2} \left|\left| \boldsymbol{B}_{i,N}^{k+1} - \boldsymbol{\mathcal{B}}_{i,N} + \boldsymbol{u_B}_{i,N}^k \right|\right|_2^2  \Bigg\}.
\end{aligned}
\end{align}

\subsubsection{Dual variable update}
\begin{align}
\boldsymbol{u_G}_{i,N}^{k+1}  =  \boldsymbol{G}_{i,N}^{k+1} - \boldsymbol{\mathcal{G}}^{k+1}_{i,N} + \boldsymbol{u_G}_{i,N}^k \label{du1}\\
\boldsymbol{u_B}_{i,N}^{k+1}  =  \boldsymbol{B}_{i,N}^{k+1}- \boldsymbol{\mathcal{B}}^{k+1}_{i,N} + \boldsymbol{u_B}_{i,N}^k \label{du2}.
\end{align}

\subsection{Implementation}
The updates of the original variable  $G_i$ and $B_i$   in \eqref{OriginalUpdatePV} and \eqref{OriginalUpdatebess} are computed in parallel. The updates of the copied variables $\mathcal{G}_i$ and $\mathcal{B}_i$ in \eqref{eq:copiedvariable} require gathering the local solutions, therefore it is solved in a centralized manner by the aggregator. 
Once the dual variables $u_{G_{i}}$ and $u_{B_{i}}$ are computed with \eqref{du1} and \eqref{du2}, they are disseminated to the subsystems together with the copied variables. 
The iterations are repeated till convergence. The convergence criterion is met when the primal residual norm ${r}^{k}$
\begin{align} 
{r}^{k+1}_{i} =  ||\boldsymbol{G}_{i,N}^{k+1} - \boldsymbol{\mathcal{G}}^{k+1}_{i,N}||_2 + ||\boldsymbol{B}_{i,N}^{k+1}- \boldsymbol{\mathcal{B}}^{k+1}_{i,N}||_2 \label{eq:primal_norm} 
\end{align}
and dual residual norm ${s}^{k}$
\begin{align}
{s}^{k+1}_{i}  = \left|\left| -\rho^{k}\Bigg[
\begin{pmatrix}
    \boldsymbol{\mathcal{B}}^{k+1}_{i,N}\\
    \boldsymbol{\mathcal{G}}^{k+1}_{i,N}
\end{pmatrix}
-
\begin{pmatrix}
    \boldsymbol{\mathcal{B}}^{k}_{i,N}\\
    \boldsymbol{\mathcal{G}}^{k}_{i,N}
\end{pmatrix} \Bigg]\right|\right|_2 \label{eq:dual_norm}
\end{align}
are both smaller than a dynamic feasibility tolerances calculated ad described in \cite{boyd2011distributed}. 

For the penalty parameter $\rho^{k}$ in \eqref{eq:adaptive_penalty}, we follow a self-adaptive scheme \cite{he2000alternating}, 
where $\tau_{incr}$ and $\tau_{decr}$ effectuate an adjustment scheme via multiplying $\rho^{k}$ by $\tau_{incr}$ when the primal residual norm is larger than the dual residual norm, and dividing $\rho^{k}$ by $\tau_{decr}$ in the opposite case\cite{boyd2011distributed}. It is to retain the primal and dual residual norms  within a factor of $\mu$ times one another as they both converge to zero. We fix $\mu$~=~10 and $\tau_{incr}$~=~2 and $\tau_{decr}$~=~2 as reported in \cite{he2000alternating}. 


\begin{align}\label{eq:adaptive_penalty}
{\rho}^{k+1} :=
\begin{cases} 
    {\tau}^{incr}{\rho}^{k}  & {r}_{i}^{k+1} > \mu {s}_{i}^{k+1}\\
    \frac{{\rho}^{k}}{{\tau}^{decr}}  & {s}_{i}^{k+1} <  \mu {r}_{i}^{k+1}\\
    {\rho}^{k}  & \text{otherwise.}
\end{cases}
\end{align}

\begin{figure}[h]
\centering
	\includegraphics[width=0.9\columnwidth]{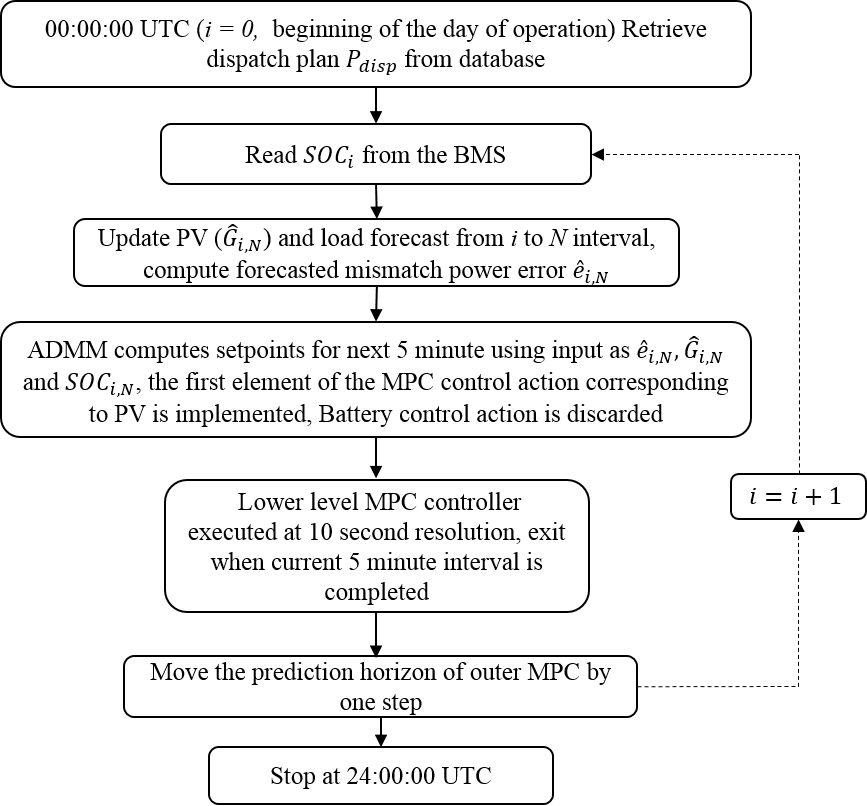}
    \caption{Flow chart showing real-time operation during 24 hours}
	\label{fig:FLOW_expt}
\end{figure}

At each computation $i$, prosumption forecast and PV generation forecasts are necessary to update the sequence $\boldsymbol{e}_{i,N}$. The former are generated with a persistent predictor, the latter are computed by applying a physical-based modelling tool-chain (transposition + PV plant models) as a function of global horizontal irradiance and air temperature predictions\footnote{provided by MeteoTest.ch.}.  The procedure describe in \cite{bright} is used to transpose
horizontal irradiance forecasts into tilted irradiance values, while the five-parameter cell model proposed in \cite{desoto}, extended to the whole PV array, is used to model the PV plant.
Also, the current battery SOC is necessary to update the initial condition of the state-of-charge integral model. The flow chart depicting the real-time operation procedure is sketched in Fig.~\ref{fig:FLOW_expt}. In the flow diagram, controllers running at two time samples: \emph{i)}~ faster MPC controller executed each 10 s and \emph{ii)}~slower ADMM based coordination mechanism computing PV set-points each 5 minute. 

The algorithm is implemented on a computer equipped with an Intel i7 processor. Statistics on the computational burden of the algorithm are discussed in the Results section.


\section{Experimental Setup}\label{Expt. setup}
The proposed control framework is validated experimentally in the real-life setup sketched in Fig.~\ref{fig:networkdiagram}, which includes office buildings with rooftop PV installations (i.e. uncontrollable prosumption), a curtailable PV facility, and a grid-connected BESS. Characteristics of these units are summarized in Table~\ref{tab:components}.
The active power flow of the distribution network is monitored by sensing the consumption at the GCP with a PMU-based metering system. The BESS is controllable by sending active/reactive power set-points to the power converter over a Modbus interface, and the PV plant is controllable by sending active power set-points to the converter over CAN communication. 

\begin{table}[!ht]
\centering
\caption{Description of connected elements at GCP}\label{tab:components}
\renewcommand{\arraystretch}{1.2}
\begin{tabular}{  m{10em} | m{10em}|  m{5em} }
   \bf{Component} & \bf{Parameter} & \bf{Value}  \\
   \hline
   \hline
   \multirow{3}{*}{Grid-connected BESS}
     & Nominal power & 720~kVA \\
     & Energy capacity & 560~kWh \\
     & Ramping rate & $\pm$20~MW/s \\
   \hline
   \multirow{3}{*}{\shortstack[l]{Prosumption (office \\ building + rooftop PV) }}
     & Peak active power demand & 350~kW \\
     & Average demand & 101~kW \\
     & Generation capacity & 82~kWp \\
   \hline
   Curtailable PV Plant & Generation capacity & 13~kWp  \\
\end{tabular}
\end{table}

\section{Experimental Results and Discussion}\label{sec:results}
\begin{figure}[!ht]
\centering
\subfloat[Dispatch plan (black), measured prosumption realization (shaded area), measured active power flow at the GCP (dashed red).]{\includegraphics[width=1\columnwidth]{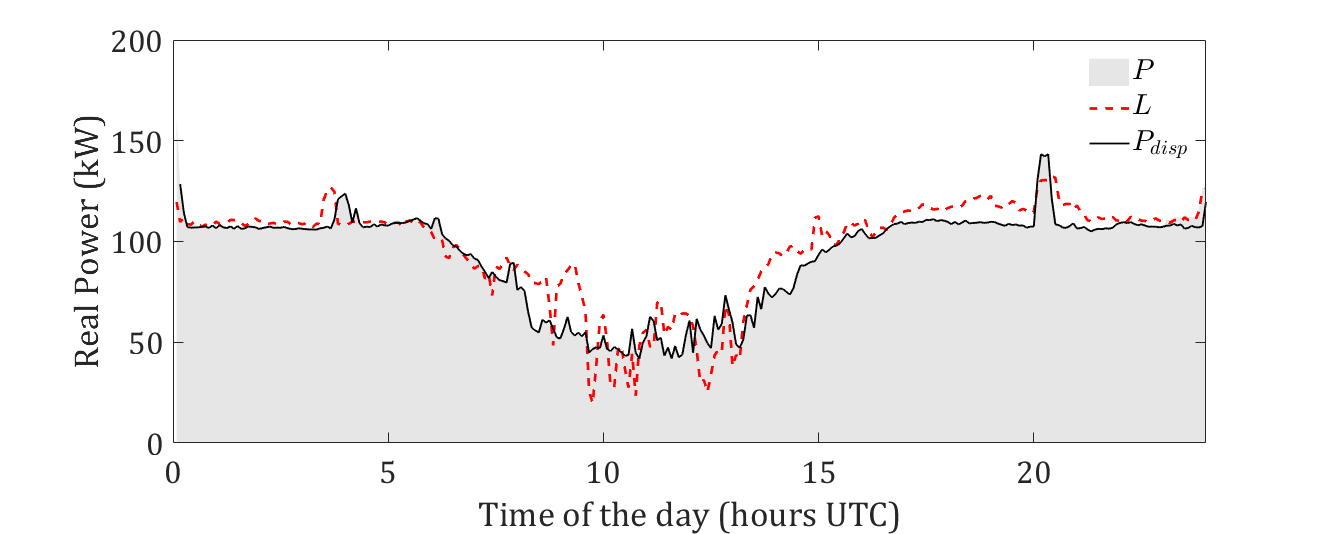}
\label{fig:Trackingplot}}
\hfil
\subfloat[Battery power injection (upper panel), and battery SOC evolution and respective limits (bottom panel).]{\includegraphics[width=1\columnwidth]{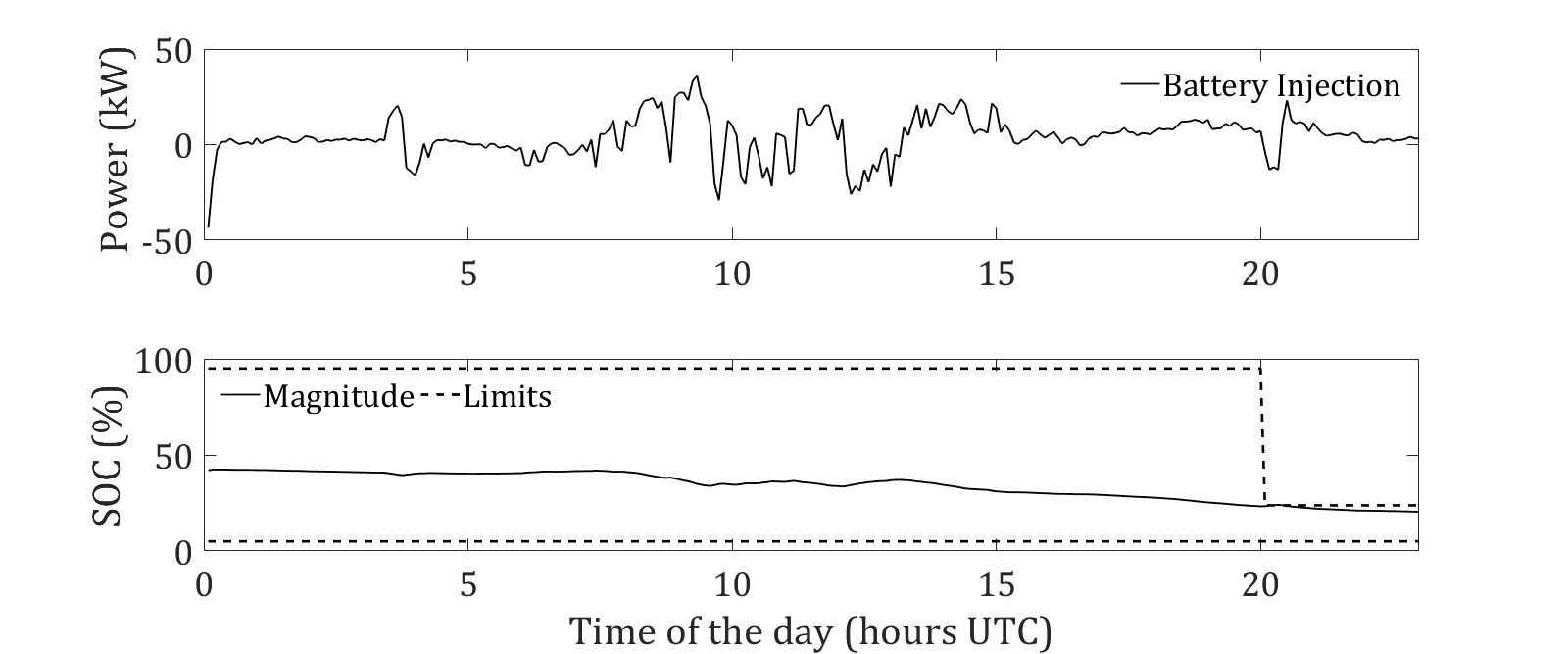}
\label{fig:SOC}}
\hfil
\subfloat[Curtailed PV (black), theoretical maximum power point (MPP) PV (dashed red) and measured PV after implementation (shaded area)]{\includegraphics[width=1\columnwidth]{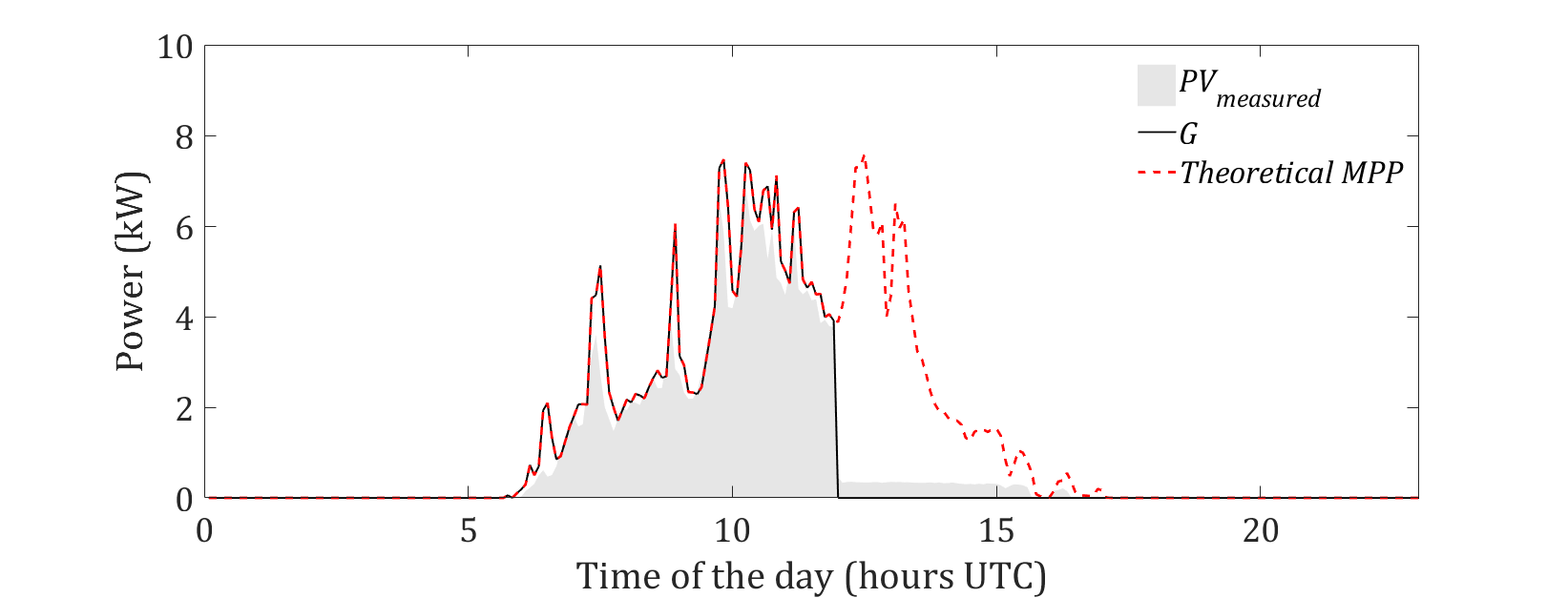}
\label{fig:PVplot}}
\caption{Operation of the dispatchable feeder with ADMM strategy on 17 September 2017.} \label{fig:experiment}
\end{figure}

The experimental results for one day of operation are shown in Fig.~\ref{fig:experiment}.
In particular, Fig.~\ref{fig:Trackingplot} shows the dispatch plan (in black), the prosumption realization (dashed red) and the realization at the GCP (gray shaded area): the prosumption realization differs from the dispatch plan due to forecasting errors, whereas the realization at the GCP, which is corrected by controlling the contribution of the battery and curtailable PV facility, achieves a good tracking of the dispatch plan.

Fig.~\ref{fig:SOC} shows the battery injection (in the upper panel), and the SOC evolution and its bounds (bottom panel, full and dashed line, respectively), which are an input of the decision problem. In this case, SOC bounds are deliberately constrained at time 20h to reproduce a situation where the flexibility is constrained (due to, e.g., the need of providing additional service or a foreseen contingency situation, like maintenance to battery modules, or simply to emulate a situation where the battery capacity is saturated). It is noteworthy that SOC evolution is always within the allowed bounds.

Fig.~\ref{fig:PVplot} shows the PV curtailment action, in particular the theoretical PV maximum power point (in red), the PV converter set-point (black) and measured PV active power measured at the converter grid connection point. As visible, during the first part of the day, the PV set-point corresponds to the maximum available power, indeed the PV plant produces close to the theoretical limit: the small differences between the two profiles are due to the fact that maximum power point operation is achieved by MPPT algorithms (maximum power point tracking), which normally rely on perturb-and-observe strategies and not on a physical model-based toolchain, as we do. As visible in Fig.~\ref{fig:PVplot}, the PV power is curtailed around midday. This is due to the distributed optimization policy, which -- as explained in the next paragraph -- decides to implement curtailment in order to satisfy the constraints of the optimization problem.

\begin{figure}[!h]
\begin{center}
\includegraphics[width=1\columnwidth]{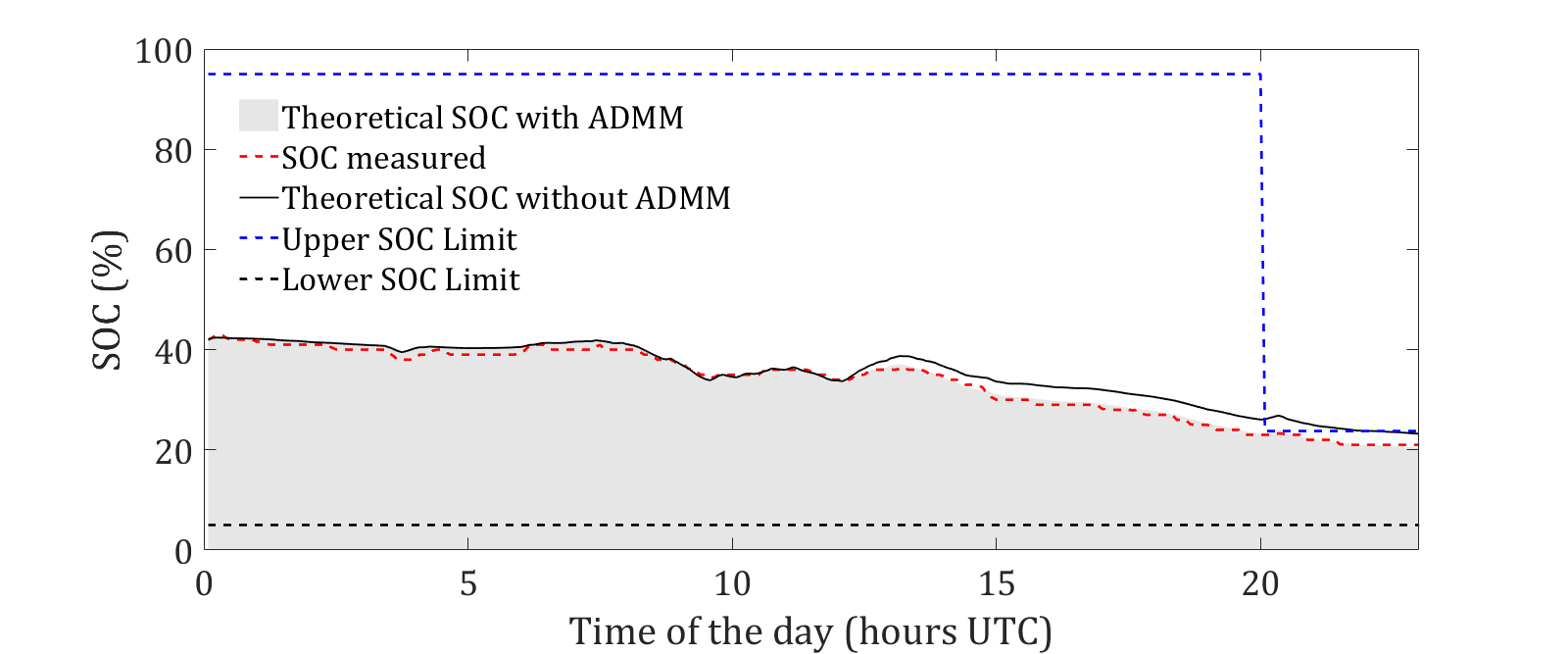}
\caption{SOC evolution with and without ADMM. The former is experimental, whereas the latter is obtained by playing back into simulations experimental data. With ADMM, SOC constraints are respected, whereas they are not without ADMM.}\label{fig:comparison}
\end{center}
\end{figure}

To evaluate whether the curtailment action determined by the distributed optimization problem was necessary, we playback the experimental measurements of the prosumption realization in a ad-hoc simulation framework\footnote{The reason why we rely on a simulation framework is the impossibility of replicating the same stochastic realization of the prosumption in two different experiments.} where the feeder dispatch is enforced by controlling the battery only. In other words, we want to verify if battery SOC constraints are still respected in the same stochastic conditions but without leveraging the controllability of the curtailable PV facility. The evolution of the SOC is modelled as in \eqref{eq:socmodel}.
Fig.~\ref{fig:comparison} compares the experimental SOC (dashed red, from the battery management system, which includes ADMM action), the simulated SOC with ADMM (shaded gray band), the simulated SOC without ADMM action, and the SOC upper and lower bounds (dashed blue and black lines). The close matching between the trajectories of the experimental and simulated SOC with ADMM denotes that the SOC estimation model \eqref{eq:soc} are in agreement and validates the reliability of the used simulation playback approach. As visible in Fig.~\ref{fig:comparison}, the experimental SOC and simulated SOC without ADMM matches until approximately midday, the time when the PV curtailment action begins (from Fig.~\ref{fig:PVplot}). After midday, the two trajectories diverge because in the latter case the battery has to charge more in order to track the dispatch plan. Nevertheless, in the former case, the battery SOC respect the SOC upper bound constraint, whereas the latter strategy (without ADMM) fails at time 20:00 UTC. We can therefore conclude that the curtailment action determined by the formulation with distributed optimization was necessary and finally accomplished the target of respecting BESS constraints. 

Numerical results comparing the control performance with and without ADMM are shown in Table~\ref{tab:cmp}. They denote that the distributed optimization control, even if it sacrifices PV generation, it achieves to respect BESS capacity constraints and that it is able to harmonize the operation of the controllable resources to achieve to dispatch stochastic prosumption.

\begin{table}[!ht]
\centering
\caption{PV generation, PV curtailments and SOC constraint violation }\label{tab:cmp}
\renewcommand{\arraystretch}{1.4}
\begin{tabular}{  m{13em} | c |  c }
   \bf{Quantity} & \bf{no ADMM} & \bf{with ADMM}  \\
   \hline
   \hline
   Max distance from SOC upper bound constraint ($>0$ violation) & 3.09~\% & -0.47~\% \\
   \hline
   PV Generation & 33.8~kWh &  21.60~kWh \\
   \hline
   PV Curtailement & -- &  12.2~kWh \\
\end{tabular}\label{curtialement_stats}
\end{table}

Table~\ref{tab:dispatch} shows the statistics on the tracking error for the case where there is no control at all (no dispatch), dispatch strategy without ADMM upper layer coordination strategy, and dispatch with ADMM. In this case, the dispatch strategy with achieves the best tracking performance with a RMS error less than 0.5~kW over 24 hours.

\begin{table}[!ht]
\centering
\caption{Tracking error statistics without dispatch, with dispatch and no ADMM, and dispatch + ADMM (kW).}\label{tab:dispatch}
\renewcommand{\arraystretch}{1.4}
\begin{tabular}{  l | c |  c | c}
   \bf{Scenario} & \bf{RMSE} & \bf{Mean} &\bf{Max} \\
   \hline
   \hline
   No dispatch & 11.1  & -4.1 & 36.0\\
   \hline
   Dispatch without ADMM & 1.60 & 0.53 & 7.8  \\
   \hline
   Dispatch + ADMM & 0.32 & $\le$ 0.01 & 2.27 \\
\end{tabular}
\end{table}

Finally, Table~\ref{tab:Iter_stats} shows mean, standard deviation and maximum value of the computation time necessary to finalize a single iteration, the number of total iteration required to convergence to a solution and the reached accuracy\footnote{Accuracy here is defined as $B_j + G_j - e_i$, i.e. the coupling constraint to achieve in \eqref{eq:cent:c1}.}. The mean time required to reach a solution is $1.01 \times 12.69 = 12.82~$seconds, denoting that the ADMM is more suitable to perform short-term energy management, as proposed in this paper, rather than real-time power control.

\begin{table}[!ht]
\centering
\caption{Computation performance} \label{tab:Iter_stats}
\renewcommand{\arraystretch}{1.4}
\begin{tabular}{  l | c |  c | c }
   \bf{Quantities} & \bf{Mean} & \bf{Standard Deviation} &\bf{Max} \\
   \hline
   \hline
   Iteration time (second) & 1.01  & 0.91 & 4.50\\
   \hline
   Iteration count & 12.69 & 2.11  & 16 \\
   \hline
   Accuracy (kW) & 0.03 &  0.14  & 1.11\\
\end{tabular}
\end{table}

\section{Conclusions and future work}\label{sec:concs}
In this paper, we proposed and experimentally validated a control and coordination framework for distributed energy resources. It consisted in two layers. The lower layer is a real-time MPC executed at 10~s resolution to achieve fine tuning of a given energy set-point. The upper layer is a slower MPC coordination mechanism based on distributed optimization and solved with ADMM. It runs each 5~minutes and it has the objective of coordinating the power flow among the resources such that enough power and energy is available in real-time to achieve a pre-established energy trajectory.

The setup was tested in a real-life setup to dispatch the operation of stochastic prosumption (e.g. office buildings with 101~kW average load, 350~kW peak demand and 82~kW peak PV generation facility) according to a dispatch plan at 5~minutes resolution, established the day before the operation. The experimental controllable elements are a 720~kVA/560~kWh battery energy storage system and a 13~kWp curtailable PV facility. 

The experimental results  showed that:
\begin{itemize}
\item the inclusion of the ADMM-based coordination mechanism successfully achieves to curtail PV generation in contingency situation (e.g., loss of battery capacity) and respect battery state-of-charge constrains;

\item the mean time to convergence to a solution in the proposed setup was 12 seconds, therefore suitable for short-term energy management;

\item imagining to cluster controllable resources into two classes (accurate and with fast actuation, like grid-connected BESS, and inaccurate and slow reaction, like PV, due to be subject to solar irradiance and MPPT dynamics), the proposed framework allows to cope with both of them and coordinate them in a cooperative and fruitful manner to track a shared control objective.

\end{itemize}
The future work is in the direction of integrating more controllable elements in the setup.

\bibliographystyle{IEEEtran}
\bibliography{faso_biblio,biblio, main}

\end{document}

%% file: fig/network_diagram.tex
\begin{tikzpicture}


\begin{scope}[shift={(0.8,0)}]
\filldraw[fill=gray!20!white, draw=black] (0.0,0.0) circle (0.20) node[anchor=west, xshift=4]{EPFL sub-transmission grid};

\begin{scope}[shift={(0,-0.75)}]
\draw[line width=0.5, -] (0,0.30) -- (0, 0.55);
\draw[draw=black] (0.0,0.0) circle (0.15);
\draw[draw=black] (0.0,0.15) circle (0.15);
\node[anchor=west] at (0.1,0.075) {Grid connection point (GCP)};
\node[anchor=east, align=center, text width=5em] at (0.1,0.075) {{\scriptsize 50/21~kV\\20~MVA}};
\draw[line width=0.5, -] (0,-0.15) -- (0,-1.1);
\draw[>=stealth, brown, <-, line width=1] (0,-0.65) -- node[anchor=west]() {{\scriptsize $P$ (composite power flow)} } (0,-0.15);
\end{scope}
\end{scope}

\begin{scope}[shift={(0,-0.8)}]
    \draw[line width=2] (0.5,-1) -- (5,-1);
    
    \draw[>=stealth, brown, ->, line width=1] (1,-1) -- node[anchor=west, yshift=5]() {{\scriptsize $L$ (aggregated feeder \emph{prosumption})} } (1.5,-1);
    
    \draw[line width=0.5, ->] (1.7,-1) -- (1.7,-1.85) node[]() {\includegraphics[width=30pt]{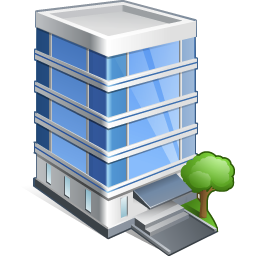}} node[xshift=-22, yshift=-28, right, text width=10em] {\small 350~kWp Prosumption (office buildings)};
    \draw[line width=0.5, ->] (2.7,-1) -- (2.7,-1.85) node[]() {\includegraphics[width=30pt]{fig/office_building.png}};
    \draw[line width=0.5, ->] (3.7,-1) -- (3.7,-1.85) node[]() {\includegraphics[width=30pt]{fig/office_building.png}};
    \draw[>=stealth, brown, <-, line width=1] (4.9,-1.40) -- node[anchor=west, text width=10em]() {{\scriptsize $G$ (PV power)} } (4.9,-2.20);
    \draw[line width=0.5, ->] (4.9,-1) -- (4.9,-2.50) node[]() {\includegraphics[width=25pt]{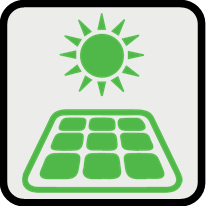}} node[xshift=-15, yshift=-28, right, text width=8em] {\small 13~kWp Curtailable PV facility};
    
    \begin{scope}[shift={(2.00,-4.90)}]
        \begin{scope}[shift={(-1.2,2.4)}]
            \draw[line width=0.5, -] (0, 1.90) -- (0, 0.30);
            \draw[draw=black] (0.0,0.0) circle (0.15);
            \draw[draw=black] (0.0,0.15) circle (0.15);
            \node[anchor=west] at (0.1,0.075) {};
            \node[anchor=east, align=center, text width=5em] at (0.1,0.075) {{\scriptsize 21/0.3~kV\\0.75~MVA}};
            \draw[line width=0.5, -] (0,-0.15) -- (0,-1.075);
            \draw[>=stealth, brown, ->, line width=1] (0,-0.5) -- node[anchor=west, yshift=-4]() {{\scriptsize $B$ (BESS charging demand)}} (0,-1.075);
        \end{scope}
        \node [anchor=south, draw=black] (battery) at (0,0) {\includegraphics[width=100pt]{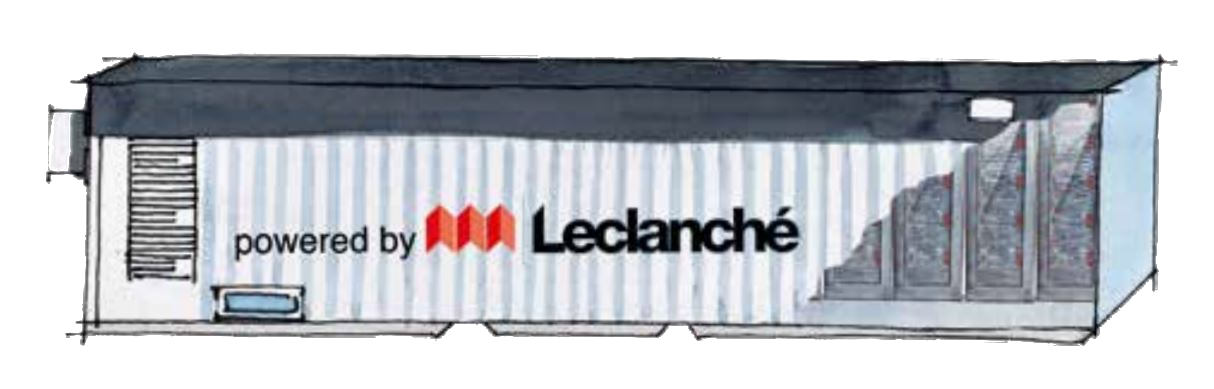}} node[anchor=north, text width=15em, align=center] {Lithium Titanate BESS\\720~kVA/500~kWh};
    \end{scope}

\end{scope}
\draw[dashed] (-1.0, -1.3) rectangle (-0.7+7.8, -0.3-6.5);

\begin{scope}[shift={(-0.7, -7.3)}]
    \draw[dashed] (0, 0) rectangle (.4, .4);
    \node[text height=1.0ex, text width=10em, anchor=west] at (0.5, 0.2){Dispatchable feeder};
    
    \draw[>=stealth, brown, ->, line width=0.7] (3.8, 0.2) -- (4.5, 0.2);
    \node[text height=1.0ex, text width=10em, anchor=west] at (4.5, 0.2){Measured power flows};
\end{scope}

\end{tikzpicture}

%% file: main.bbl
\begin{thebibliography}{10}
\providecommand{\url}[1]{#1}
\csname url@samestyle\endcsname
\providecommand{\newblock}{\relax}
\providecommand{\bibinfo}[2]{#2}
\providecommand{\BIBentrySTDinterwordspacing}{\spaceskip=0pt\relax}
\providecommand{\BIBentryALTinterwordstretchfactor}{4}
\providecommand{\BIBentryALTinterwordspacing}{\spaceskip=\fontdimen2\font plus
\BIBentryALTinterwordstretchfactor\fontdimen3\font minus
  \fontdimen4\font\relax}
\providecommand{\BIBforeignlanguage}[2]{{%
\expandafter\ifx\csname l@#1\endcsname\relax
\typeout{** WARNING: IEEEtran.bst: No hyphenation pattern has been}%
\typeout{** loaded for the language `#1'. Using the pattern for}%
\typeout{** the default language instead.}%
\else
\language=\csname l@#1\endcsname
\fi
#2}}
\providecommand{\BIBdecl}{\relax}
\BIBdecl

\bibitem{6598997}
K.~Christakou, D.-C. Tomozei, J.-Y. Le~Boudec, and M.~Paolone, ``Gecn: Primary
  voltage control for active distribution networks via real-time
  demand-response,'' \emph{IEEE Transactions on Smart Grid}, vol.~PP, no.~99,
  2013.

\bibitem{vovos2007centralized}
P.~N. Vovos, A.~E. Kiprakis, A.~R. Wallace, and G.~P. Harrison, ``Centralized
  and distributed voltage control: Impact on distributed generation
  penetration,'' \emph{IEEE Transactions on Power Systems}, vol.~22, no.~1, pp.
  476--483, 2007.

\bibitem{hu2014coordinated}
J.~Hu, S.~You, M.~Lind, and J.~Ostergaard, ``Coordinated charging of electric
  vehicles for congestion prevention in the distribution grid,'' \emph{IEEE
  Transactions on Smart Grid}, vol.~5, no.~2, pp. 703--711, 2014.

\bibitem{6558529}
P.~Douglass, R.~Garcia-Valle, P.~Nyeng, J.~Ostergaard, and M.~Togeby, ``Smart
  demand for frequency regulation: Experimental results,'' \emph{IEEE
  Transactions on Smart Grid,}, vol.~4, no.~3, pp. 1713--1720, Sept 2013.

\bibitem{guerrero2013advanced}
J.~M. Guerrero, M.~Chandorkar, T.-L. Lee, and P.~C. Loh, ``Advanced control
  architectures for intelligent microgrids—part i: Decentralized and
  hierarchical control,'' \emph{IEEE Transactions on Industrial Electronics},
  vol.~60, no.~4, pp. 1254--1262, 2013.

\bibitem{molina2011decentralized}
A.~Molina-Garcia, F.~Bouffard, and D.~S. Kirschen, ``Decentralized demand-side
  contribution to primary frequency control,'' \emph{IEEE Transactions on Power
  Systems}, vol.~26, no.~1, pp. 411--419, 2011.

\bibitem{liu2014decentralized}
W.~Liu, W.~Gu, W.~Sheng, X.~Meng, Z.~Wu, and W.~Chen, ``Decentralized
  multi-agent system-based cooperative frequency control for autonomous
  microgrids with communication constraints,'' \emph{IEEE Transactions on
  Sustainable Energy}, vol.~5, no.~2, pp. 446--456, 2014.

\bibitem{Hammerstrom2007}
D.~J. Hammerstrom, R.~Ambrosio, T.~A. Carlon, J.~G. Desteese, R.~Kajfasz, and
  R.~G. Pratt, ``{Pacific Northwest GridWise Testbed Demonstration Projects
  Part I . Olympic Peninsula Project},'' \emph{Contract}, p. 157, 2007.

\bibitem{8013070}
E.~Dall’Anese, S.~Guggilam, A.~Simonetto, Y.~C. Chen, and S.~V. Dhople,
  ``Optimal regulation of virtual power plants,'' \emph{IEEE Transactions on
  Power Systems}, vol.~PP, no.~99, pp. 1--1, 2017.

\bibitem{Bernstein2015}
A.~Bernstein, L.~Reyes-Chamorro, J.-Y.~L. Boudec, and M.~Paolone, ``A
  composable method for real-time control of active distribution networks with
  explicit power setpoints. part i: Framework,'' \emph{Electric Power Systems
  Research}, 2015.

\bibitem{7542590}
F.~Sossan, E.~Namor, R.~Cherkaoui, and M.~Paolone, ``Achieving the
  dispatchability of distribution feeders through prosumers data driven
  forecasting and model predictive control of electrochemical storage,''
  \emph{IEEE Transactions on Sustainable Energy}, vol.~7, no.~4, pp.
  1762--1777, Oct 2016.

\bibitem{6913566}
M.~Abu~Abdullah, K.~Muttaqi, D.~Sutanto, and A.~Agalgaonkar, ``An effective
  power dispatch control strategy to improve generation schedulability and
  supply reliability of a wind farm using a battery energy storage system,''
  \emph{Sustainable Energy, IEEE Transactions on}, vol.~6, 2015.

\bibitem{fabietti2017j}
L.~Fabietti, T.~T. Gorecki, E.~Namor, F.~Sossan, M.~Paolone, and C.~N. Jones,
  ``Enhancing the dispatchability of distribution networks through electric
  energy storage systems and flexible demand: Control architecture and
  experimental validation,'' \emph{Submitted to Energy and Buildings,
  Elsevier}, 2017.

\bibitem{saele2011demand}
H.~S{\ae}le and O.~S. Grande, ``Demand response from household customers:
  Experiences from a pilot study in norway,'' \emph{IEEE Transactions on Smart
  Grid}, vol.~2, no.~1, pp. 102--109, 2011.

\bibitem{borenstein2002dynamic}
S.~Borenstein, M.~Jaske, and A.~Rosenfeld, ``Dynamic pricing, advanced
  metering, and demand response in electricity markets,'' 2002.

\bibitem{sundstrom2012flexible}
O.~Sundstrom and C.~Binding, ``Flexible charging optimization for electric
  vehicles considering distribution grid constraints,'' \emph{IEEE Transactions
  on Smart Grid}, vol.~3, pp. 26--37, 2012.

\bibitem{7948761}
E.~Stai, L.~Reyes-Chamorro, F.~Sossan, J.~Y.~L. Boudec, and M.~Paolone,
  ``Dispatching stochastic heterogeneous resources accounting for grid and
  battery losses,'' \emph{IEEE Transactions on Smart Grid}, vol.~PP, no.~99,
  pp. 1--1, 2017.

\bibitem{erseghe2014distributed}
T.~Erseghe, ``Distributed optimal power flow using admm,'' \emph{IEEE
  transactions on power systems}, vol.~29, no.~5, pp. 2370--2380, 2014.

\bibitem{dall2014decentralized}
E.~Dall’Anese, S.~V. Dhople, B.~B. Johnson, and G.~B. Giannakis,
  ``Decentralized optimal dispatch of photovoltaic inverters in residential
  distribution systems,'' \emph{IEEE Transactions on Energy Conversion},
  vol.~29, no.~4, pp. 957--967, 2014.

\bibitem{tan2014optimal}
Z.~Tan, P.~Yang, and A.~Nehorai, ``An optimal and distributed demand response
  strategy with electric vehicles in the smart grid,'' \emph{IEEE Transactions
  on Smart Grid}, vol.~5, no.~2, pp. 861--869, 2014.

\bibitem{giselsson2017linear}
P.~Giselsson and S.~Boyd, ``Linear convergence and metric selection for
  douglas-rachford splitting and admm,'' \emph{IEEE Transactions on Automatic
  Control}, vol.~62, no.~2, pp. 532--544, 2017.

\bibitem{7419888}
R.~Halvgaard, L.~Vandenberghe, N.~K. Poulsen, H.~Madsen, and J.~B. Jørgensen,
  ``Distributed model predictive control for smart energy systems,'' \emph{IEEE
  Transactions on Smart Grid}, vol.~7, no.~3, pp. 1675--1682, May 2016.

\bibitem{boyd2011distributed}
S.~Boyd, N.~Parikh, E.~Chu, B.~Peleato, and J.~Eckstein, ``Distributed
  optimization and statistical learning via the alternating direction method of
  multipliers,'' \emph{Foundations and Trends in Machine Learning}, vol.~3,
  2011.

\bibitem{scolari_2018}
E.~Scolari, L.~Reyes, F.~Sossan, and M.~Paolone, ``A comprehensive assessment
  of the short-term uncertainty of grid-connected pv systems,'' \emph{IEEE
  Transactions on Sustainable Energy}, vol.~PP, no.~99, pp. 1--1, 2018.

\bibitem{scolari2017}
E.~Scolari, F.~Sossan, and M.~Paolone, ``Photovoltaic model-based solar
  irradiance estimators: Performance comparison and application to maximum
  power forecasting,'' \emph{IEEE Transactions on Sustainable Energy}, 2017.

\bibitem{he2000alternating}
B.~He, H.~Yang, and S.~Wang, ``Alternating direction method with self-adaptive
  penalty parameters for monotone variational inequalities,'' \emph{Journal of
  Optimization Theory and applications}, vol. 106, no.~2, pp. 337--356, 2000.

\bibitem{bright}
J.~Bright, C.~Smith, P.~Taylor, and R.~Crook, ``Stochastic generation of
  synthetic minutely irradiance time series derived from mean hourly weather
  observation data,'' \emph{Solar Energy}, vol. 115, no. Supplement C, pp. 229
  -- 242, 2015.

\bibitem{desoto}
W.~D. Soto, S.~Klein, and W.~Beckman, ``Improvement and validation of a model
  for photovoltaic array performance,'' \emph{Solar Energy}, vol.~80, no.~1,
  pp. 78 -- 88, 2006.

\end{thebibliography}
